\documentclass[aps,prd,preprint,nofootinbib]{revtex4}
\usepackage{amsfonts}
\parskip + 5pt
\usepackage{epsfig}
\usepackage{graphicx}
\usepackage{float}

\begin{document}
\title{ $J/\psi$ Pair Production at the Tevatron with
$\sqrt{s}=1.96~\mathrm{TeV}$\\[7mm]}

\author{Cong-Feng Qiao$^{1,2}$ and Li-Ping Sun$^{1}$}
\affiliation{$^{1}$College of Physical Sciences, Graduate University
of Chinese Academy of Sciences \\ YuQuan Road 19A, Beijing 100049,
China} \affiliation{$^{2}$Theoretical Physics Center for Science
Facilities (TPCSF), CAS\\ YuQuan Road 19B, Beijing 100049, China}

\author{~\vspace{0.7cm}}

\begin{abstract}

We study the $J/\psi$ pair production issue at the Fermilab Tevatron
Run II with the center-of-mass energy $\sqrt{s}=1.96~\mathrm{TeV}$.
Both the color-singlet and -octet production mechanisms are
considered. Our result shows that the transverse momentum($p_T$)
scaling behaviors of double $J/\psi$ differential cross sections in
color-singlet and -octet deviate distinctively from each other while
$p_T$ is larger than $8~ \mathrm{GeV}$, and with the luminosity of
${5 \mathrm{fb^{-1}}}$ the $J/\psi$ pair events from color-singlet
scheme are substantially measurable in Tevatron experiments, even
with certain lower transverse momentum cut. Hence the Tevatron is
still possibly a platform to check the heavy quarkonium
production mechanism.\\


\end{abstract}

\maketitle

\begin{figure}[b,m,u]
\centering
\includegraphics[width=1.000\textwidth]{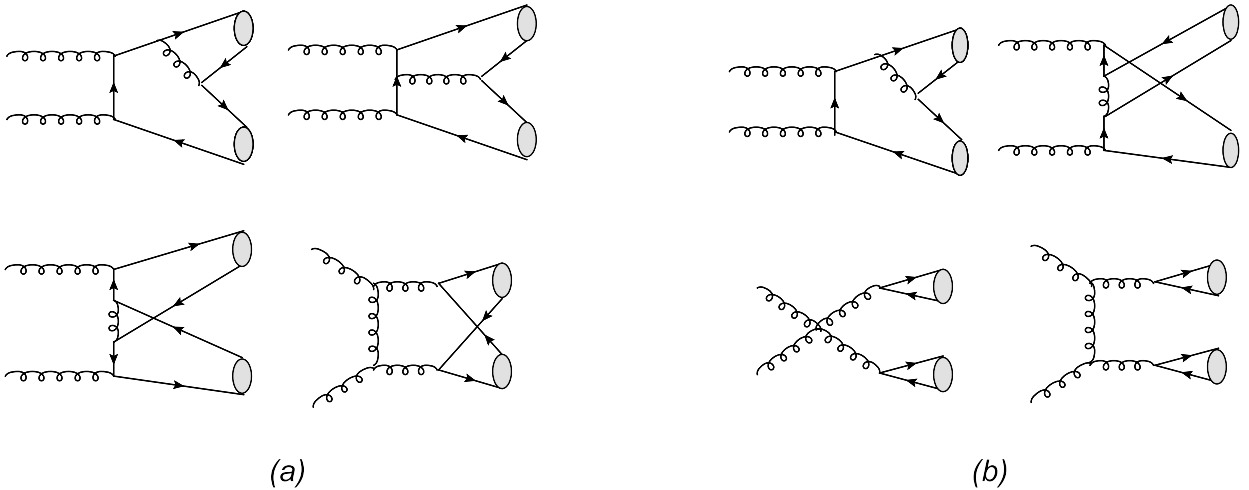}%
\caption{\small The typical Feynman diagrams of $J/\psi$ pair
production at leading order. Figure (a) belongs to the CS scheme,
while Figure (b) is the CO case .} \label{graph1}
\end{figure}

The Tevatron \cite{tev1,tev2} Run II with the center-of-mass energy
$\sqrt{s}=1.96~\mathrm{TeV}$ is a good platform for the study of the
heavy quark mesons. In this brief report, we reevaluate the $J/\psi$
pair production rate at the Tevatron in the framework of
non-relativistic chromodynamics (NRQCD) \cite{nrqcd}. In the
color-singlet model, the partonic subprocesses start at order
$\alpha_s^4$, which include $g + g \rightarrow J/\psi + J/\psi$ and
$q + \bar{q} \rightarrow J/\psi + J/\psi$. Intuitively, the latter,
the quark-antiquark annihilation process, contributes less than the
former at the Tevatron, hence in the following analysis we mainly
focus on the gluon-gluon process as shown in Figure \ref{graph1}.

The differential cross section for $J/\psi$ pair hadroproduction
reads
\begin{eqnarray}
\label{eq:3} \frac{d\sigma}{d p_T} (p p \rightarrow 2J/\psi + X)  =
\sum_{a,b}\int dy_1 dy_2 f_{a/p}(x_a) f_{b/p}(x_b) 2 p_T x_a x_b
\frac{d\hat{\sigma}} {d{t}} (a + b \rightarrow 2J/\psi)\; ,
\end{eqnarray}
where $f_{a/p}$ and $f_{b/p}$ denote the parton densities in proton
or antiproton; $y_1$, $y_2$ are the rapidity of the two produced
$J/\psi$s. The partonic scattering process, the gluon-gluon to
polarized and unpolarized $J/\psi$ pair differential cross section
$\frac{d\hat{\sigma}} {d{t}}$, can be calculated in the standard
method, which had been performed before in
Refs.\cite{singlet1,singlet2}, and we confirm the analytic result.

In NRQCD, the color-octet scheme is guaranteed \cite{com}. For the
$J/\psi$ pair production process, the typical Feynman diagrams are
shown in Figure \ref{graph1}. Part of it, the lower two ones in CO
mechanism, the fragmentation processes, in the figure was evaluated
in Ref.\cite{vbarger}. In this work, we consider not only the
$J/\psi$ pair in configuration of
$|c\bar{c}[^3S_1^{(8)}]gg\rangle|c\bar{c}[^3S_1^{(8)}]gg\rangle$,
but also in configuration of
$|c\bar{c}[^3S_1]\rangle|c\bar{c}[^3S_1^{(8)}]gg\rangle$, though the
latter contributes less in the end.

Except for the difference in CO and CS non-perturbative matrix
element projections, the perturbative calculations of Feynman
diagrams for both CS and CO are similar. In numerical calculation,
we enforce the Tevatron experimental condition, the pseudorapidity
cut $|\eta(J/\psi)| < 2.0$, the central-of-mass energy
$\sqrt{S}=1.96~\mathrm{TeV}$ for Tevatron Run II. The input
parameters take the values \cite{tev2}
\begin{eqnarray}
m_c =  1.5\; \rm{GeV},\; |R(0)|^2 = 0.8\;\rm{GeV}^3,\; <{\cal
O}^{J/\psi}_8({}^3S_1)> = 0.012\;\rm{GeV}^3\;.\label{eq:2}
\end{eqnarray}

With the above formulas and inputs, one can readily obtain the
polarized $J/\psi$ pair production cross section at the Tevatron. In
the numerical calculation, the parton distribution of CTEQ5L
\cite{cteq} is used. The integrated cross section $\sigma(p\;
{p}\rightarrow J/\psi J/\psi)$ with various $p_T$ lower bounds is
presented in Table~\ref{ratio1} for the CS and CO production schemes
respectively, where the the branching fraction of $ B(J/\psi \to
\mu^+ \mu^-)= 0.0597$ is taken into account.

\begin{table}
\begin{center}
\caption{The integrated cross sections of $J/\psi$ pair production
under various low transverse momentum cuts. Here, $\bot\bot$
represents the situation in which both $J/\psi$s qre transversely
polarized, $\|\|$ represents that both $J/\psi$s are longitudinally
polarized, $\|\bot$ represents that one $J/\psi$ is longitudinally
polarized and the other is transversely polarized. The $tot_{18}$ in
the last row represents the double $J/\psi$ yields from CS + CO
production scheme for reference. }\small \vspace{1mm}
\begin{tabular}{|c||c|c|c|c|c|c|c|c|c|c|c|c|c|c|c|}
\hline\hline &\multicolumn{5}{|c|}{CS Model}&\multicolumn{5}{|c|}{CO
Model}\\
\hline\hline $\sigma\setminus p_{Tcut}$&~3 GeV~ & ~4 GeV~ & ~5 GeV~
& ~6 GeV~ & ~7 GeV~ &~3 GeV~
& ~4 GeV~ & ~5 GeV~ & ~6 GeV~ & ~7 GeV~\\
\hline\hline $\bot\bot$ & 0.520pb & 0.145pb & 0.044pb & 0.015pb &
5.408fb & 0.047pb & 0.033pb
& 0.021pb & 0.014pb & 8.869fb\\
\hline\hline $\|\|$ &0.214pb &0.074pb &0.025pb &8.927fb &3.411fb
&0.345fb &0.102fb &0.032fb
 &0.011fb &0.004fb\\
\hline\hline $\|\bot$ & 0.547pb & 0.131pb & 0.032pb & 8.424fb
&2.466fb &5.303fb &2.640fb &1.289fb
 &0.636fb &0.323fb\\
\hline\hline $tot$ & 1.278pb & 0.348pb & 0.101pb & 0.032pb &0.011pb
&0.053pb &0.035pb &0.023pb
 &0.014pb &9.195fb\\
\hline\hline $tot_{18}$ & -- & -- & -- & -- &-- &0.040pb &0.011pb
&3.384fb
 &1.107fb &0.400fb\\
\hline\hline
\end{tabular}
\label{ratio1}
\end{center}
\end{table}

\begin{figure}[H]
\includegraphics[width=0.500\textwidth]{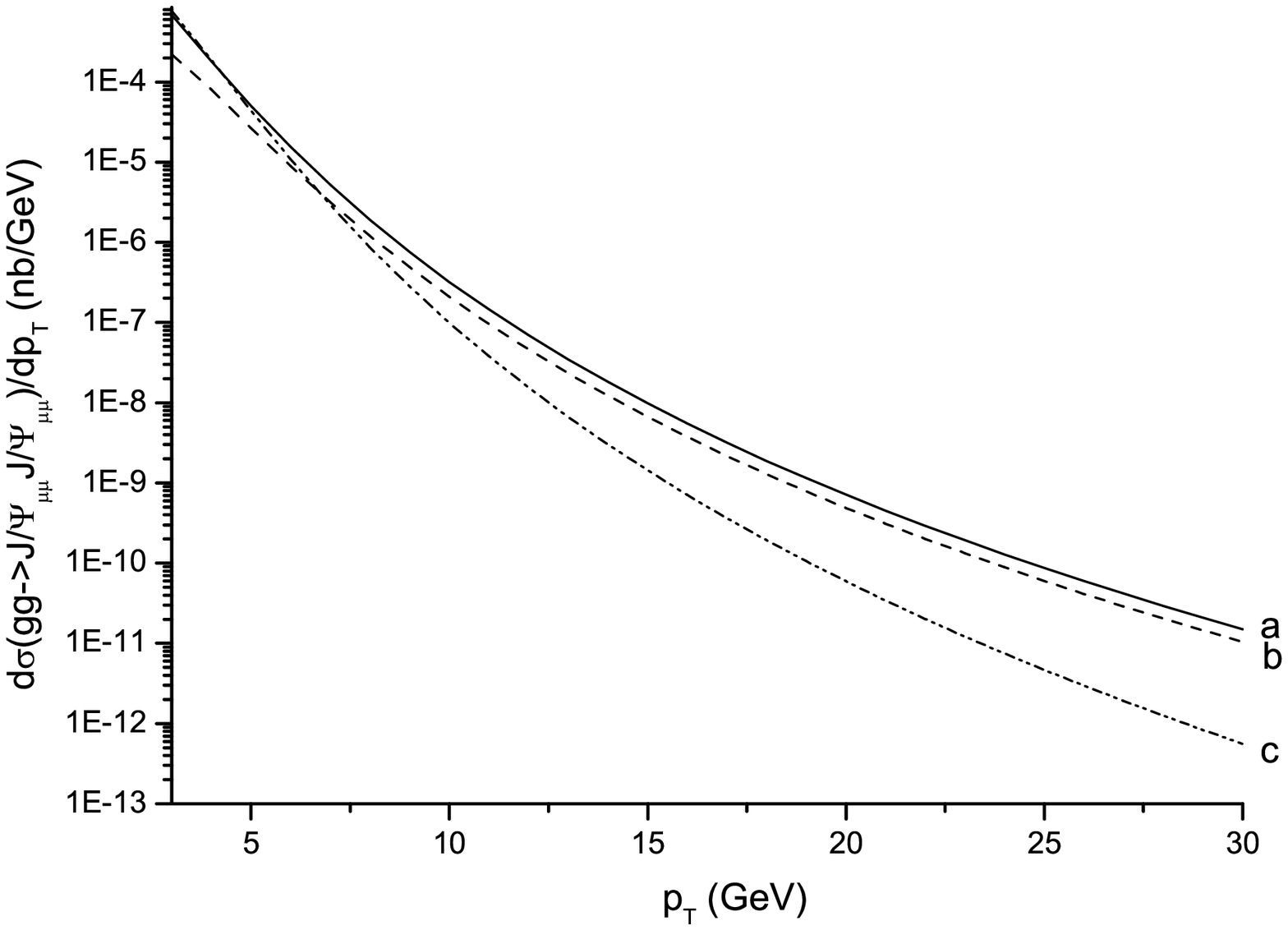}\hspace*{\fill}
\hspace{-8mm}
\includegraphics[width=0.500\textwidth]{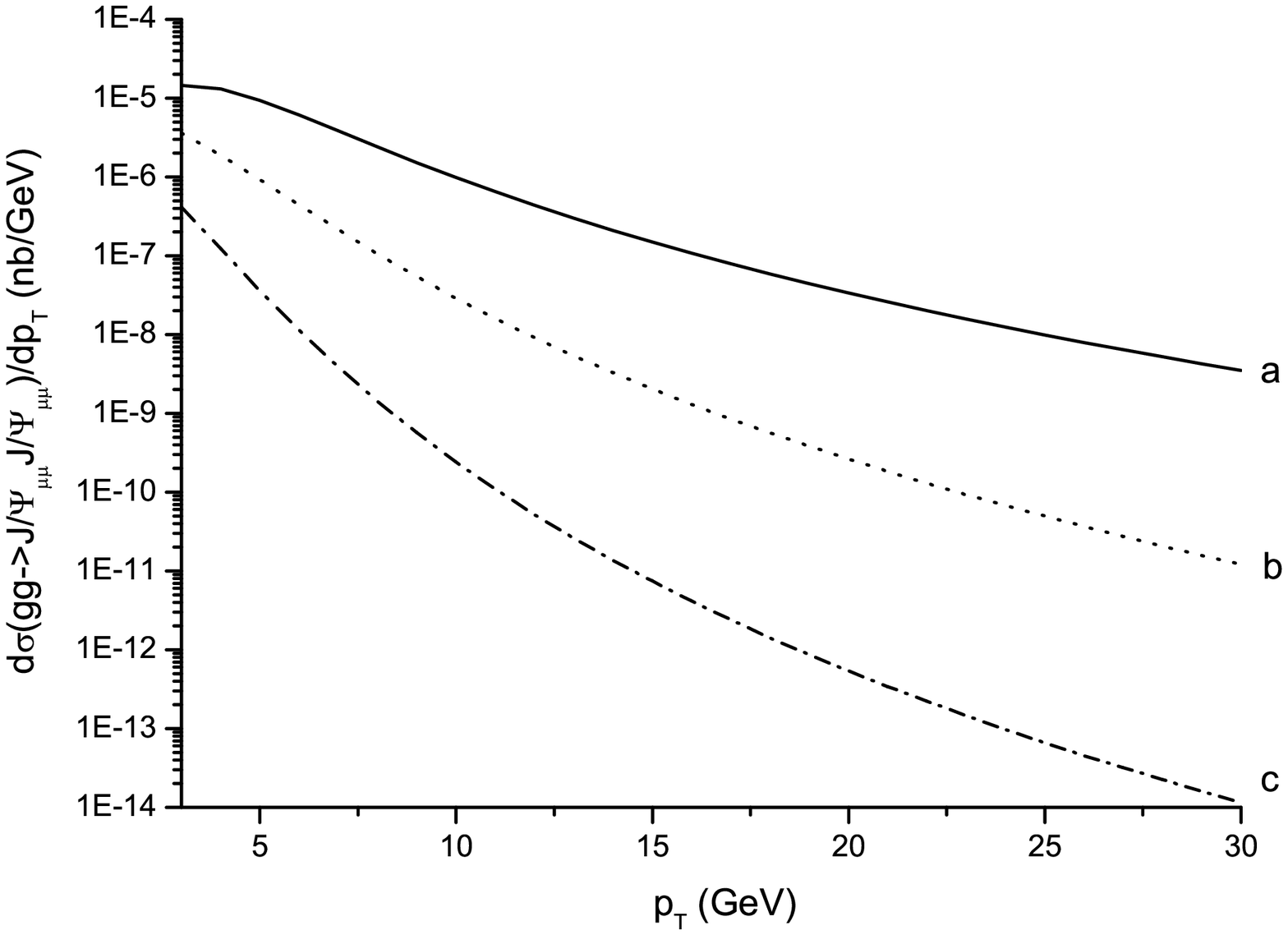}\hspace*{\fill}
\hspace{-8mm} \caption{\small The differential cross-section of
$J/\psi$ pair production versus $p_T$ at the Tevatron. The left
figure represents the yields from color-singlet, the lines from top
to bottom, i.e. a ,b and c, denote $\bot\bot$, $\|\|$, and $\|\bot$
cases, respectively. The right figure represents the yields from
color-octet, the lines from top to bottom, i.e. a, b and c, denote
$\bot\bot$, $\|\bot$, and $\|\|$ cases, respectively.} \label{lpty1}
\vspace{-0mm}
\end{figure}

\begin{figure}[H]
\centering
\includegraphics[width=0.600\textwidth]{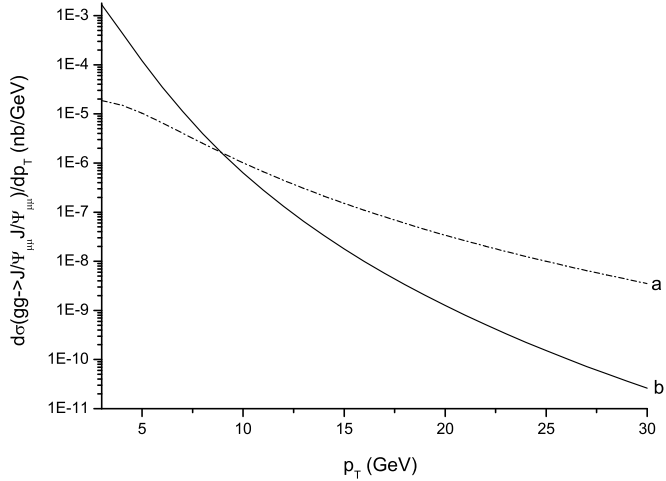}
\caption{\small The differential cross-section of $J/\psi$ pair
production versus $p_T$ at the Tevatron. Lines a and b, represent
the yields from color-octet and color-singlet in unpolarized case,
respectively.} \label{lpty2} \vspace{-0mm}
\end{figure}

The spectra of double-$J/\psi$ exclusive production as a function of
transverse momentum $p_T$ are illustrated in Figures \ref{lpty1} and
\ref{lpty2}. Figure \ref{lpty1} shows that at large $p_T$, in the CS
scheme the contribution from $\bot\bot$ case dominates the process,
while in the CO case, the $\bot\bot$ dominates the process in all
$p_T$ region. Figure \ref{lpty2} indicates that the conventional CS
production scheme dominates over the CO one in relatively low-$p_T$
region, while $p_T < 8~\mathrm{GeV}$.

In conclusion, we evaluate the $J/\psi$ pair production at the
Fermilab Tevatron Run II energy. With the luminosity of
$\sim{5~\mathrm{fb^{-1}}}$, we find that there is a large number of
$J/\psi$ pair events produced there. Imposing a low transverse
momentum cut of 7 GeV, the observed data should come only from the
CO mechanism, and the detection efficiency be 10\% or lower. In all,
it is very hopeful to observe the $J/\psi$ pair production process
in the Tevatron II experiments, and even to check the charmonium
production mechanism through it.

\vskip 0.7cm \noindent {\bf Acknowledgments}

We would like to thank Dmitry Bandurin of D0 Collaboration for
bringing our attention to this issue. This work was supported in
part by the National Natural Science Foundation of China(NSFC) under
Grant No. 10935012, 10821063 and 11175249.



\begin{thebibliography}{99}

\bibitem{tev1} E. Braaten, S. Fleming and A.K. Leibovich \ Phys.\
 Rev.\ D{\bf 63}, 094006(2001).

\bibitem{tev2} P. Cho and Leibovich, Phy. Rev.\ D{\bf 53}, 150(1996);
 {\it ibid}, 6203(1996).

\bibitem{nrqcd} G.T. Bodwin, E. Braaten, and G.P. Lepage, Phys.\
 Rev.\ D{\bf 51}, 1125(1995).

\bibitem{singlet1} Cong-Feng Qiao, Phys. Rev. D{\bf 66}, 057504(2002).

\bibitem{singlet2} Cong-Feng Qiao, Li-Ping Sun, and Peng Sun,
J. Phys.\ G{\bf 37}, 075019(2010).

\bibitem{com} G.T. Bodwin, E. Braaten, and G.P. Lepage, Phys.\ Rev.\
 D{\bf 46}, R3703(1992); E. Braaten and S. Fleming, Phys.\ Rev.\
 Lett.\ {\bf 74},  3327(1995).

\bibitem{vbarger} V. Barger, S. Fleming, and R.J.N. Phillips, Phys.\
Lett.\ B{\bf 371}, 111(1996).

\bibitem{cteq} CTEQ Collaboration, H.L. Lai {\it et al.}, Eur.\ Phys.\ J.\
 C{\bf 12}, 375(2000).

\end{thebibliography}
\end{document}